\documentclass[twocolumn,showpacs,aps,pra]{revtex4-1}
\usepackage[english]{babel}
\usepackage[latin1]{inputenc}
\usepackage{hyperref}
\usepackage{amsmath, amssymb, amsfonts, mathrsfs}
\usepackage{graphicx}
\graphicspath{{images/}}
\usepackage{amssymb,xcolor}

\usepackage{xcolor}
\usepackage{exscale}
\usepackage{graphicx}
\usepackage{amsmath}
\usepackage{latexsym}
\usepackage{amsfonts}
\usepackage{amssymb}

\newcommand{\beq}{\begin{equation}}
\newcommand{\eeq}{\end{equation}}

\newcommand{\proj}[1]{|#1\rangle\langle#1|}

\newcommand{\id}{\leavevmode\hbox{\small1\normalsize\kern-.33em1}}

\DeclareMathOperator{\Tr}{Tr}
\DeclareMathOperator{\Id}{ {\bf 1}}

\def\pmx{\begin{pmatrix}}
\def\emx{\end{pmatrix}}

\newcommand{\map}[1]{\mathscr{#1}}

\newcommand{\cnot}{\text{CNOT}}

\newcommand{\bra}[1]{\left\langle{#1}\right\vert}
\newcommand{\ket}[1]{\left\vert{#1}\right\rangle}

\begin{document}

\title{Quantum channel detection}

\author{C. Macchiavello and M. Rossi}
\affiliation{Dipartimento di Fisica and INFN-Sezione di Pavia, 
via Bassi 6, 27100 Pavia, Italy}

\date{\today}
\begin{abstract}
We present a method to detect properties of quantum channels, assuming
that some a priori information about the form of the channel is available. 
The method is based on a correspondence with entanglement detection methods
for multipartite density matrices based on witness operators.
We  first illustrate the method in the case of entanglement breaking 
channels and non separable random unitary channels, and show how it can be 
implemented experimentally by means of local measurements. We then study the detection 
of non separable maps and show that for pairs of systems of dimension 
higher than two the detection operators are not the same as in the random unitary case, highlighting a richer separability structure of quantum channels with respect to quantum states. Finally we consider the set of PPT maps, developing a technique to reveal NPT maps.

\end{abstract}

\maketitle

\section{introduction}

The possibility of determining properties of quantum communication channels
or quantum devices is of great importance in order to be able to design and
operate the channel at the best of its performances. In many realistic implementations some a priori information  on the form of a quantum channel, or a quantum noise process, is available and it is of large interest to 
determine experimentally whether or not the channel
has a certain property. The aim of this work is to propose efficient 
methods 
to detect this possibility by avoiding full quantum process tomography, 
which allows a complete reconstruction of the channel but it requires a large 
number of measurement settings.  At the same time, from the 
point of view of implementations, our procedure is experimentally feasible 
with present day technology based on local measurements.

This work is organized as follows. In Sec. \ref{math} we will explain our main idea, treating as an introductory example entanglement breaking channels. In Sec. \ref{sru} and \ref{sep} we will study the cases of separable random unitaries and separable maps, respectively. We will develop a method to detect NPT channels in Sec. \ref{ppt} and we summarize the main results in Sec. \ref{conc}.

\section{main idea and entanglement breaking channels}
\label{math}

In this section we will show the main idea of the proposed quantum channel detection method and its link to entanglement detection methods for multipartite quantum systems. To this aim we remind that quantum channels, and in general quantum noise processes, 
are described by completely positive (CP) and trace preserving (TP) maps $\map{M}$, which can be expressed in the Kraus form \cite{kraus}
as
\begin{equation}
\map{M}[\rho]=\sum_k A_k\rho A_k^\dagger,
\end{equation}
where $\rho$ is the density operator of the quantum system on which the 
channel acts and the Kraus operators $\{A_k\}$ fulfil the TP constraint 
$\sum_k A_k^\dagger A_k=\Id$. 

\begin{figure}[t!]
\includegraphics[width=\columnwidth]{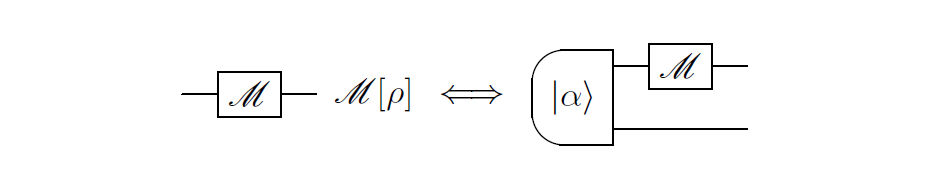}
\caption{Scheme showing the Choi-Jamiolkowski isomorphism: 
on the left the map $\map{M}$, on the right the corresponding Choi state 
$C_\map{M}$.}
\label{scheme}
\end{figure}

The detection method proposed is based on the use of the 
Choi-Jamiolkowski isomorphism \cite{jam}, which gives a one-to-one
correspondence between CP-TP maps acting on $\mathcal{D(H)}$ (the set of 
density operators on $\mathcal{H}$, with arbitrary finite dimension $d$) 
and bipartite density operators 
$C_{\map{M}}$ on $\mathcal{H\otimes H}$  with $\Tr_A[C_{\map{M}}]=\Id_B/d$. This isomorphism can be described as follows
\begin{equation}
\map{M}\Longleftrightarrow 
C_{\map{M}}=(\map{M}\otimes\map{I})[\ket{\alpha}\bra{\alpha}],
\end{equation}
where $\map{I}$ is the identity map, 
and $\ket{\alpha}$ is the maximally entangled state with respect to the 
bipartite space $\mathcal{H\otimes H}$, i.e. $\ket{\alpha}=\frac{1}{\sqrt 
d}\sum_{k=1}^d\ket{k}\ket{k}$. This is schematically depicted in Fig.
\ref{scheme}.
In this work, by the above isomorphism, we link some specific properties 
of quantum channels to properties of the corresponding Choi states 
$C_{\map{M}}$. 
We will consider properties that are based on a convex structure of the 
quantum channels. 

Consider as a first simple case 
the class of entanglement breaking (EB) channels 
\cite{EB}. A possible definition for an EB channel is
based on the separability of its Choi state: a quantum 
channel is EB if and only if its Choi state is separable.
This allows to formulate a method to detect whether a quantum channel is not
EB by exploiting entanglement detection methods designed
for bipartite systems \cite{ent-wit}. To this end, we remind the concept of
entanglement detection via witness operators \cite{horo-ter}: 
a state $\rho$ is entangled 
if and only if there exists a hermitian operator $W$ such that 
$\Tr[W\rho]< 0$ and $\Tr[W\rho_{sep}]\geq 0$ for all separable states.

As a simple example of quantum channel detection consider the case of 
qubits and the single qubit depolarising channel, defined as
\begin{equation}\label{depo}
\Gamma_{\{p\}}[\rho]= \sum^3_{i=0}{p_i \sigma_i \rho \sigma_i},
\end{equation}
where $\sigma_0$ is the identity operator, \{$\sigma_i$\} ($i=1,2,3$) 
are the three Pauli operators $\sigma_x ,\sigma_y, \sigma_z$ respectively
(for brevity of notation in the following the Pauli operators will be
denoted by $X$, $Y$ and $Z$), 
and $p_0=1-p$ (with $p\in[0,1]$), while $p_i=p/3$ for $i=1,2,3$.
Such a channel is EB for $p \geq 1/2$. 
The corresponding set of Choi bipartite density operators 
is given by the Werner states
\begin{equation}\label{werner states}
\rho_p= (1-\frac{4}{3}p)\proj{\alpha}
+\frac{p}{3}\Id\;.
\end{equation}
It is then possible to detect whether a depolarising channel is not 
entanglement 
breaking by exploiting an entanglement witness operator for the above set
of states \cite{ent-wit,jmo}, which has the form
\begin{equation}\label{wEB}
W_{EB}= \frac{1}{4}(\Id\otimes\Id 
-X\otimes X + Y\otimes Y -Z\otimes Z)\;.
\end{equation}
The method can then be implemented by preparing a two-qubit state
in the maximally entangled state  $\ket{\alpha}$, then operating with the 
quantum channel to be detected on one of the two qubits and measuring the 
operator $W_{EB}$ acting on both qubits at the end.
If the resulting average value is negative, we can then conclude that the channel under 
consideration is not EB.

We will now prove that our method provides also a lower bound on a particular feature of EB channels recently defined in Ref. \cite{giova} as follows. Let $\map{M}$ be a generic map acting on a $d$-dimensional system and $\map{D}_\sigma$ the completely depolarizing channel defined as $\map{D}_\sigma[\rho]=\sigma$, where $\sigma$ is an arbitrary state. The quantity $\mu_c(\map{M})$ is defined as the minimum value of the mixing probability parameter $\mu\in [0,1]$ that transforms the convex combination
$(1-\mu) \map{M} + \mu \map{D}_\sigma$ into an entanglement breaking (EB) channel, i.e. in formulae
\begin{equation}\label{mu}
 \mu_c(\map{M})=\min_{\sigma} \left\{\mu|(1-\mu) \map{M} + \mu\map{D}_\sigma \in \mbox{EB}\right\}.
 \end{equation}
By the  Choi-Jamiolkowski isomorphism, we can rephrase the definition \eqref{mu} in term of Choi states as
\begin{equation}\label{mubis}
 \mu_c(\map{M})=\min_{\sigma} \left\{\mu|(1-\mu) C_\map{M} + \mu \sigma  \otimes  \frac{\Id}{d} \in Sep \right\},
\end{equation}
and link this quantity to the well-known generalized robustness of entanglement. Given a state $\rho$, the generalized robustness of entanglement is defined \cite{VidalTarrach, Steiner} as the minimal $s>0$ such that the state
$\frac{\rho + s \sigma}{1 + s}$ is separable, where $\sigma$ is an arbitrary state (not necessarily separable), namely
\begin{equation}\label{robust}
R(\rho)=\min_\sigma\{s|\frac{\rho + s \sigma}{1 + s}\in Sep \}.
\end{equation}
This quantity can be interpreted as the minimum amount of noise necessary to wash out completely the entanglement initially present  in the state $\rho$. Thus, by defining $p_c(\rho)=1-\frac{1}{1+R(\rho)}$ and interpreting $\rho$ as the Choi state $C_\map{M}$ corresponding to the map $\map{M}$, we can bound $\mu_c(\map{M})$ as
\begin{equation}
\mu_c(\map{M})\geq p_c(C_\map{M}),
\end{equation}
since the minimising set involved in the definition \eqref{mubis} of $\mu_c(\map{M})$ is smaller than the minimising set considered for $R(C_\map{M})$, as can be seen in Eq. \eqref{robust}. By the above inequality we can derive a bound for the generalized robustness from the experimental data of an entanglement detection procedure \cite{Brandao1} as
\begin{equation}\label{bound}
R(\rho)\geq  |c|/w_{max},
\end{equation}
where $c$ is measured experimentally via the expectation value of the witness, i.e. $\Tr[W\rho]=c<0$, while  $w_{max}$ is the maximal eigenvalue of the operator $W$. As a result, we can find that
\begin{equation}
\mu_c(\map{M})\geq 1-\frac{1}{1+|c|/w_{max}},
\end{equation}
which links the expectation value of the witness measured experimentally to the theoretical quantity $\mu_c(\map{M})$. In the case of the depolarising channel \eqref{depo} with $p<1/2$, by using the witness $W_{EB}$ given by Eq. \eqref{wEB}, the above bound takes the form
\begin{equation}
\mu_c(\Gamma_{\{p\}})\geq \frac{1-2p}{2-2p}.
\end{equation}
In this case, however, the bound is not tight since the theoretical $\mu_c(\Gamma_{\{p\}}) $ can be computed to be $\frac{2-4p}{3-4p}$ by following the method developed in \cite{giova}.

\section{Separable random unitaries}
\label{sru}

We will now consider the case of random 
unitary (RU) channels, defined as 
\begin{equation}\label{RU}
\map{U}[\rho]=\sum_k p_k U_k\rho U_k^\dagger,
\end{equation}
where $U_k$ are unitary operators and $p_k> 0$ with $\sum_k p_k=1$. Notice 
that this kind of maps includes several interesting models of quantum noisy
channels, such as the already mentioned depolarising channel or 
the phase damping channel and the bit flip channel \cite{NC}. 
RUs were also studied 
extensively and characterised in Ref. \cite{scheel}.

We will now consider the case where the RU channel acts
on a bipartite system $\rho_{AB}$ as follows
\begin{equation}\label{SRU}
\map{V}[\rho_{AB}]=\sum_k p_k (V_{k,A}\otimes W_{k,B})\rho_{AB} (V_{k,A}^\dagger
\otimes W_{k,B}^\dagger),
\end{equation}
where both $V_{k,A}$ and $W_{k,B}$  
are unitary operators for all $k$'s, acting on systems A and B respectively. 
Quantum channels of the
above form are named separable random unitaries (SRUs) and they form a convex
subset in the set of all CP-TP maps acting on bipartite systems. 
Interesting examples of channels of this form are given by Pauli memory 
channels \cite{memory}.
\begin{figure}[t!]
\includegraphics[width=\columnwidth]{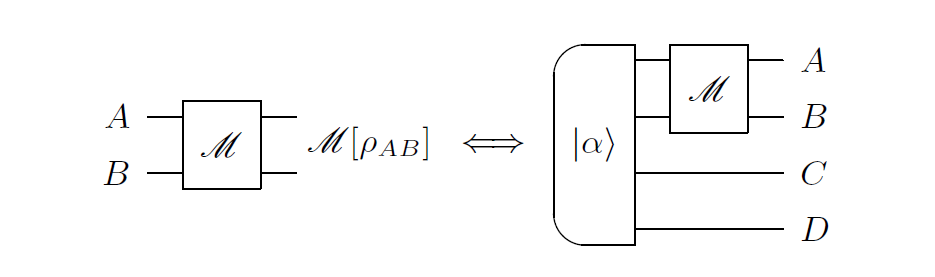}
\caption{Scheme of the Choi-Jamiolkowski isomorphism in the 
case of four-partite states. The state $\ket{\alpha}$ on the right is the 
maximally entangled state with respect to the bipartition AB|CD.}
\label{scheme2}
\end{figure}

The Choi state corresponding to quantum channels acting on bipartite systems
is a four-partite state (composed of systems A, B, C and D),
as shown in Fig. \ref{scheme2}.
Notice that the state 
$\ket{\alpha}=\frac{1}{\sqrt{d_{AB}}}\sum_{k,j=1}^{d_{AB}}
\ket{k,j}_{AB}\ket{k,j}_{CD}$ (where $d_{AB}=d_Ad_B$ 
is now the dimension of the 
Hilbert space of the bipartite system AB) can also be written as  
$\ket{\alpha}=\ket{\alpha}_{AC}
\ket{\alpha}_{BD}$, namely it is a biseparable state for the partition AC|BD
of the global four-partite system.
The Choi states corresponding to SRU channels therefore form a convex set, 
which is a subset of all biseparable states for the partition AC|BD.
Since the generating set of SRUs is
given by local unitaries $U_A\otimes U_B$,  
the generating biseparable pure states in the corresponding set of Choi states 
have the form
\begin{equation}\label{ext}
\ket{U_A\otimes U_B}=(U_A\otimes\Id_{C})\ket{\alpha}_{AC}\otimes(U_B\otimes\Id_{D})
\ket{\alpha}_{BD}\;.
\end{equation}
We name the set of four-partite Choi states corresponding to SRUs as $S_{SRU}$. It is now possible to design detection procedures for SRU maps by employing suitable
witness operators that detect the corresponding Choi state with
respect to biseparable states (in AC|BD) belonging to $S_{SRU}$. 

We will now focus on the case of a unitary transformation 
$U$
acting on two $d$-dimensional systems. The corresponding Choi state is pure
and has the form
\begin{equation}\label{U}
\ket{U}=(U\otimes\Id)\ket{\alpha}
\end{equation}
Therefore, a suitable detection operator for $U$ as a non SRU gate
can be constructed as 
\begin{equation}\label{W}
W_{SRU,U}=\alpha_{SRU}^2\Id-C_U\;,
\end{equation}
where $C_U=\ket{U}\bra{U}$, and the coefficient $\alpha_{SRU}$ is the overlap between the closest biseparable state in the set $S_{SRU}$ and the entangled state $\ket{U}$, namely
\begin{equation}
\alpha_{SRU}^2 =\max_{\map{M}_{SRU}}\bra{U}C_{\map{M}_{SRU}}\ket{U}.
\end{equation}
Notice that, since the maximum of a linear function over a convex set is 
always achieved on the extremal points, the maximum above can be always 
calculated by maximising over the pure biseparable states (\ref{ext})
\cite{notabisep}, i.e.
\begin{equation}
\alpha_{SRU} =\max_{U_A,U_B}|\bra{U_A\otimes U_B}U\rangle |=\frac{1}{d^2}\max_{U_A,U_B}|\Tr[(U_A^\dagger\otimes U_B^\dagger)U]|.
\end{equation}

\begin{figure}[t!]
\includegraphics[width=\columnwidth]{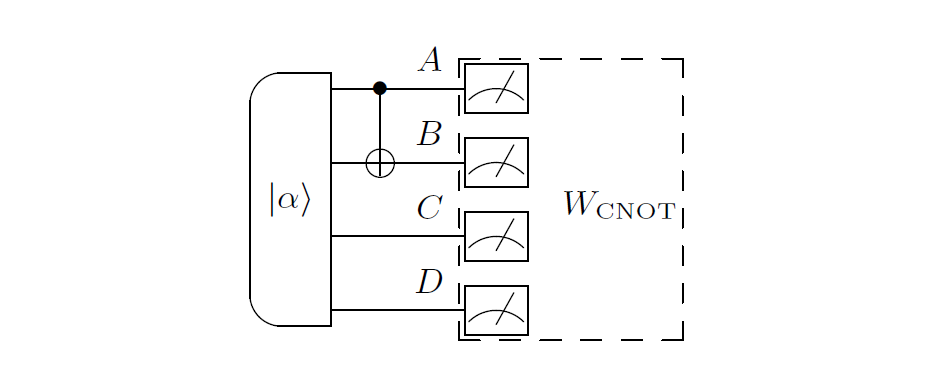}
\caption{Experimental scheme implementing the detection of the $\cnot$ gate.}
\label{scheme-exp}
\end{figure}

As an example of the above procedure consider the CNOT gate 
acting on a two-qubit system, defined by
\begin{equation}
\cnot=
\begin{pmatrix}
\Id & 0 \\
0 & X 
\end{pmatrix}\;,
\end{equation}
with $\Id$ representing the $2\times 2$ identity matrix, 
and $X$ the usual Pauli operator. The coefficient $\alpha_{SRU}$ for $U=\cnot$  can be computed as follows. 
The state \eqref{U} specialized for the CNOT gate is clearly not separable with respect to the split AC|BD and it can be expressed in the Schmidt decomposition regarding that split as 
\begin{equation}
\ket{\cnot}=\frac{1}{\sqrt 2}(\ket{00}_{AC}\ket{\alpha}_{BD}
+\ket{11}_{AC}\ket{\psi^+}_{BD})\;,
\end{equation}
where $\ket{\psi^+}=\frac{1}{\sqrt 2}(\ket{01}+\ket{10})$.
The above expression naturally proves that the maximum overlap with any 
biseparable state w.r.t. AC|BD cannot exceed the value of $1/\sqrt 2$. 
Since the convex set $S_{SRU}$ of allowed states in our optimisation problem 
is smaller than the set of all biseparable states, this would give us only an 
upper bound for the maximum overlap $\alpha_{SRU}$. However, two local unitary 
operations $U_A$ and $U_B$ that saturate this bound can be explicitly 
found, namely $U_A=S$ and $U_B=e^{-i\frac{\pi}{4}X}$,
where $S$ is the phase gate given by $S=diag(1,i)$. 
This finally proves that the optimal coefficient $\alpha_{SRU}$ equals $1/\sqrt 2$ even
if we restrict to the set of biseparable states $S_{SRU}$. 
Moreover, the detection operator $W_\cnot=\frac{1}{2}\Id - C_\cnot$ can be decomposed 
into a linear combination of local operators as follows
\begin{align}\label{WCNOT}
W_\cnot =\frac{1}{64}(&31\Id\Id\Id\Id -\Id X\Id X-XXX\Id -X\Id XX \nonumber \\
				&-ZZ\Id Z+ZY\Id Y+YYXZ+YZXY\nonumber \\
				&-Z\Id Z\Id -ZXZX+YXY\Id +Y\Id YX\nonumber \\
				&-\Id ZZZ+\Id YZY + XYYZ+XZYY)\;,
\end{align}
where for simplicity of notation the tensor product symbol has been omitted. As we can see from the above form, the CNOT can be detected by using 
nine different local measurements settings, namely $\{XXXX, ZZZZ, ZYZY, YXYX,
YYXZ, YZXY, \\ ZXZX, XYYZ, XZYY\}$.
Actually, in the first line of the above expression the expectation values of operators $\Id X\Id X,XXX\Id,X\Id XX $ can be obtained by 
measuring the operator $XXXX$ and suitably processing the experimental data.
Similar groupings can be done for the other terms in (\ref{WCNOT}), such that
the only measurement settings needed are the nine listed above.
Following \cite{ent-wit, gu-hy},  it can be also easily proved that the 
above form is optimal in the sense that 
it involves the smallest number of  measurement settings.
From an experimental point of view, the optimal detection procedure can be implemented
as follows: prepare a four-partite qubit system in the state  
$\ket{\alpha}=\ket{\alpha}_{AC}\ket{\alpha}_{BD}$, apply 
the quantum channel to qubits A and B, and finally perform the set of nine local measurements
reported above in order to measure the operator (\ref{WCNOT}). 
If the resulting average value is negative then the quantum channel is 
detected as a non SRU map.
The experimental scheme is shown in Fig. \ref{scheme-exp}.  
Notice that the number of measurements needed in this procedure is much smaller 
than the one required for complete quantum process tomography, since the former scales as $d_{AB}^2$ \cite{jmo} while the latter as $d_{AB}^4$ \cite{NC}.

The number of measurement settings in the detection scheme can be further 
decreased if we allow a non optimal detection operator, in the sense that the 
coefficient $\alpha_{SRU}$ in $W_\cnot$ is smaller than the maximum value. 
In this case, since the state $C_{\cnot}$ is a stabilizer state with generators 
$\{XXX\Id,\Id X\Id X, Z\Id Z\Id, ZZ\Id Z\}$, an alternative detection
operator can be derived, following the approach of Ref. \cite{tg}. The 
resulting suboptimal detection operator turns out to be
\begin{align}\label{WCNOTsub}
\tilde W_\cnot=3\Id -2&\left[\frac{(\Id+XXX\Id)}{2}
\frac{(\Id+\Id X\Id X)}{2}\right. \nonumber\\
&+\left.\frac{(\Id+ Z\Id Z\Id)}{2}\frac{(\Id +ZZ\Id Z)}{2}\right],
\end{align}
which requires only the two local measurement settings $\{XXXX,ZZZZ\}$.
The robustness of the method in the detection of the CNOT gate
was analysed in \cite{sinaia}.

\section{Separable maps}
\label{sep}

We will now focus on the detection of 
non separable maps.
By definition, a separable map $\map{M}_{sep}$ is given by
\begin{equation}
\map{M}_{sep}[\rho_{AB}]=\sum_k (A_k\otimes B_k)\rho_{AB} (A_k^\dagger\otimes B_k^\dagger),
\end{equation}
namely it can be written in terms of separable Kraus operators \cite{sep1}. Here we do not require the TP condition. Notice that the set of separable maps is a larger set than  the set of SRUs studied above.
A general map $\map{M}$ acting on two qudits is not separable if and only 
if the corresponding Choi state $C_\map{M}$ is entangled with respect to the 
splitting  AC|BD \cite{kraus_sep}. 

Analogously to the case of SRU maps, for a unitary transformation $U$ we can 
define a witness operator of the same form \eqref{W}, where now the coefficient $\alpha_{SRU}^2$ is replaced by $\alpha_{S}^2$ defined as
\begin{equation}\label{alpha_S}
\alpha_{S}^2=\max_{\map{M}_{sep}}\bra{U}C_{\map{M}_{sep}}\ket{U}.
\end{equation}
Since the set of SRUs is a subset of all separable maps, in  general 
$\alpha_S \ge \alpha_{SRU}$. 
The maximum in Eq. \eqref{alpha_S} is attained on pure states, which are the
extremal points in the set of $C_{\map{M}_{sep}}$. 
Since a map $\map{M}$ is described by a single Kraus operator 
if and only if its Choi state $C_\map{M}$ is pure \cite{chi?}, we 
can then compute the maximum on separable maps $\map{M}_{sep}$ with 
a single Kraus operator. The calculation for $\alpha_{S}$ can then be simplified as
\begin{equation}\label{overlap}
\alpha_{S}=\max_{A, B}|\bra{A\otimes B}U\rangle|=\frac{1}{d^2}\max_{A, B}|\Tr[(A^\dagger\otimes B^\dagger)U]|.
\end{equation}
Notice that now we do not require $A\otimes B$ to be TP, otherwise both $A$ 
and $B$  would be automatically unitary. 
Interestingly, we will now show that for a general unitary $U$ on two-qubit
systems the two coefficients $\alpha_{SRU}$ and $\alpha_{S}$ coincide, while 
for higher dimension this does no longer hold.

We will compute the coefficients by starting from the Schmidt decomposition 
of an operator $O$ acting on two qudits, which can be written as
\begin{equation}\label{sd}
O=\sum_{i=1}^r \sigma_i A_i\otimes B_i,
\end{equation}
where $\{A_i\}_{i=1,...,d^2}$ and $\{B_i\}_{i=1,...,d^2}$ are two orthogonal bases ($\Tr[A_i^\dagger A_j]=\Tr[B_i^\dagger B_j]=d\delta_{ij}$) for the operator space, and $r$ is the Schmidt rank fulfilling $1\leq r\leq d^2$. Notice that
the unique Schmidt coefficients $\sigma_i$ are always positive and ordered, i.e. $\sigma_1\geq\dots\geq\sigma_r$. As a result, if we write the unitary $U$ in the Schmidt decomposition \eqref{sd}, it follows that the maximum 
\eqref{overlap} is achieved by the choice of $A\otimes B= A_1\otimes B_1$, where $A_1$ and $B_1$ are the operators corresponding to the 
largest Schmidt coefficient $\sigma_1$. We then have
\begin{equation}\label{overlap_single}
\alpha_{S}=\frac{1}{d^2}|\Tr[(A_1^\dagger\otimes B_1^\dagger) U]|=\sigma_1.
\end{equation}
It is then interesting to establish whether the optimal separable operator 
$A_1\otimes B_1$ has to be unitary as well. As mentioned above, we will 
show that this is true for qubit systems but does no longer hold when the 
dimension increases.
We will first show that for two qubits it is always possible to 
find a separable unitary $U_A\otimes U_B$ such that the overlap with 
$U$ achieves the maximum $\sigma_1$, namely 
\begin{equation}\label{statem2}
\exists\text{ } U_A,U_B \text{ s.t. } |\bra{U_A\otimes U_B} U\rangle |
=\alpha_{SRU}=\sigma_1.
\end{equation}
This is a consequence of the Cartan decomposition 
\cite{kraus_cart,nielsen_cart} of a general unitary $U$ acting on two qubits,
given by
\begin{equation}\label{cartan}
U=(V_A\otimes V_B) \tilde U (W_A\otimes W_B),
\end{equation}
where $V_A,V_B,W_A$ and $W_B$ are single qubit unitaries and
\begin{equation}
\tilde U=e^{i(\theta_x X\otimes X+\theta_y Y\otimes Y+\theta_z Z\otimes Z)}.
\end{equation}
 
Notice that, by the definitions $c_\alpha=\cos \theta_\alpha$ 
and $s_\alpha=\sin \theta_\alpha$, $\tilde U$ takes the form
\begin{align}
\tilde U=(c_xc_yc_z+is_xs_ys_z)\Id\otimes\Id+(c_xs_ys_z+is_xc_yc_z)X\otimes X \nonumber\\
+(s_xc_ys_z+ic_xs_yc_z)Y\otimes Y+(s_xs_yc_z+ic_xc_ys_z)Z\otimes Z. 
\end{align}
According to \eqref{cartan}, it is then straightforward to see that the above form of $\tilde U$ directly leads to the Schmidt decomposition of $U$. 
Actually, the magnitudes of the coefficients in front of the bipartite 
operators correspond to the Schmidt coefficients themselves and the phases 
can be reabsorbed into the Pauli operators without changing the 
orthogonality relations. Therefore, given a unitary $U$ on two qubits, 
it is always possible to find a local unitary achieving the maximum 
$\sigma_1$, since there always exists a Schmidt decomposition of $U$ 
involving only unitary operators as local basis. For higher dimensional 
systems the above argument does not hold.
Actually, already in the two-qutrit case it may happen that the maximum 
\eqref{overlap_single} can, in general, be attained only by local non unitary operators. This  means that the closest (under the criterion defined in \eqref{overlap}) separable map to a unitary $U$ may be non unitary. 

We show an explicit example for a system of two qutrits given by 
the gate $Z_3$ defined as
\begin{equation}
Z_3=\text{diag}(1,1,1,1,1,1,1,1,-1),
\end{equation}
which is unitary and not separable. 
We can rewrite $Z_3$ in the Schmidt form with Schmidt rank $r=2$ as
\begin{equation}
Z_3=\sigma_1 A_1\otimes B_1 +\sigma_2 A_2\otimes B_2,
\end{equation}
where $\sigma_{1,2}=\sqrt{\frac{1}{2}(9\pm \sqrt{17})}/3$, while the 
operators $A_{1,2}$ and $B_{1,2}$ are non unitary and can be written as
\begin{align}
A_{1,2}=\frac{\sqrt 3}{\sqrt{102\pm 22\sqrt{17}}}\times & \label{A12}\\
				\times\text{diag}(5&\pm\sqrt{17},5\pm\sqrt{17},1\pm \sqrt{17}),\nonumber \\
B_{1,2}=\frac{\sqrt 3}{\sqrt{646\pm 150\sqrt{17}}}\times & \label{B12}\\
			\times\text{diag}(11&\pm 3\sqrt{17},11\pm 3\sqrt{17},9\pm \sqrt{17}).\nonumber
\end{align}
From the Schmidt decomposition it immediately follows that the value of the  
maximum overlap is given by 
$\alpha_S=\sigma_1=\sqrt{\frac{1}{2}(9+ \sqrt{17})}/3\sim 0.854$. 
The coefficient $\alpha_{SRU}$ can be computed, 
leading to $\alpha_{SRU}\sim 0.786$ \cite{nota-3}. 
Hence, this proves that the maximum attained over SRUs is always strictly 
smaller then the maximum achieved by separable maps, 
$\alpha_{SRU}<\alpha_{S}$. 
We want to stress that our method is then suitable to detect the gap between 
separable and SRU maps, as long as $d\ge 3$. 
Actually, by the amount of violation of the expectation value of $W_{SRU,U}$ for 
detecting $U$, we can establish whether the detected map was 
separable or in addition random unitary too. For example the unitary $Z_3$ can be detected as a non SRU map by a witness operator of the following form
\begin{equation}
W_{SRU,Z_3}=\alpha_{SRU}^2\Id - C_{Z_3},
\end{equation}
where $C_{Z_3}=\ket{Z_3}\bra{Z_3}$ and $\alpha_{SRU}\sim 0.786$. 
Moreover the expectation value of $W_{SRU,Z_3}$ 
over the Choi state of the experimentally accessible map $\map M$, i.e. 
$\Tr[W_{SRU,Z_3}C_\map{M}]$, allows us to distinguish between non SRU and non 
separable maps. Actually, $\map M$ is detected to be non SRU if 
$\Tr[W_{SRU,Z_3}C_\map{M}]<0$, and in addition we can say
that $\map M$ is not a 
separable map if $\Tr[W_{SRU,Z_3}C_\map{M}]<\alpha_{SRU}^2-\alpha_S^2$. 

\section{PPT channels}
\label{ppt}

In this section we will consider a larger set of quantum channels, namely PPT channels. A CP map $\map{M}$ acting on two qudits is positive partial transpose (PPT) if and only if the composite map $\map{M_T}=\map{T}_A \circ\map{M}\circ\map{T}_A$, being $\map{T}_A$ the partial transposition map on the first system $A$, is CP \cite{rains, npt}. 
Since a map $\map M$ is CP if and only if the corresponding Choi operator 
$C_{\map M}$ is positive, we can restate the above definition as: a CP map 
$\map M$ is PPT if and only if the Choi operator $C_{\map{M_T}}$ related to the composite map $\map{M_T}$ is positive. 

By the above correspondence we will develop a method to detect 
whether a map is non-positive partial transpose (NPT). 
We will employ techniques already developed for the detection of 
entangled NPT states \cite{copositive}, 
namely we consider a witness operator of the following form
\begin{equation}\label{npt}
W_{\text{PPT}}=\ket{\lambda_{-}}\bra{\lambda_{-}}^{\map{T}_A},
\end{equation}
where $\ket{\lambda_{-}}$ is the eigenvector of the Choi state
$C_{\map{M_T}}$ corresponding to the most negative eigenvalue $\lambda_{-}$ 
for an NPT map 
$\map{M}$. 

\begin{figure}[t!]
\includegraphics[width=\columnwidth]{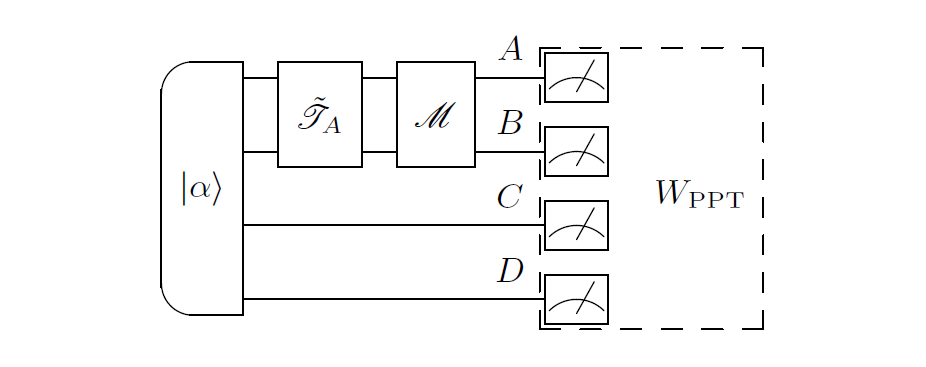}
\caption{Experimentally-feasible scheme to implement the detection 
of the NPT map $\map{M}$.}
\label{nonphy}
\end{figure}

The expectation value of the above witness operator should now be measured 
for the Choi operator corresponding to the composite map 
$\map{M}\circ\map{T}_A$, since the partial transposition following $\map{M}$ is already taken into account in the form of the operator (\ref{npt}).
Therefore, a crucial point of this approach is now related to  
the implementation of the map $\map{T}_A$, which is non CP. 
A possible solution is to add noise to the 
map $\map{T}_A$ in order to make it CP, as shown in Ref. 
\cite{ekert}. Following the approach of \cite{ekert}
we consider the minimal amount of depolarising noise
such that the following map
\begin{equation}\label{noisy}
\tilde{\map{T}_A}[\rho_{AB}]=(1-p)\map{T}_A[\rho_{AB}]+ p\frac{\Id_{AB}}{d^2}
\end{equation}
is CP. This is given by $p=d^3/(d^3+1)$ \cite{ekert}.
From an experimental point of view, we then consider the implementation of
the map $\tilde{\map{T}_A}$ instead of the non-physical map $\map{T}_A$, 
as shown in Fig. \ref{nonphy}. 
This procedure will lead to an extra contribution in the expectation
value of the witness operator, related to the 
presence of the depolarized term in Eq. \eqref{noisy}. 
The expectation value of $W_{\text{PPT}}$ for the Choi state
$C_{\map{M}\circ\tilde{\map{T}_A}}$ related to the composite map
$\map{M}\circ\tilde{\map{T}_A}$ is given by
\begin{align}\label{diff}
&\Tr[ W_{\text{PPT}}C_{\map{M}\circ\tilde{\map{T}_A}}] = \\
&	=(1-p)\bra{\lambda_-}C_{\map{M_T}}\ket{\lambda_-}
			+p \bra{\lambda_-}\map{M_T}[\frac{\Id_{AB}}{d^2}]\otimes\frac{\Id_{CD}}							{d^2}\ket{\lambda_-} \nonumber\\
&	=(1-p)\lambda_- +p \bra{\lambda_-}\map{M_T}[\frac{\Id_{AB}}{d^2}]\otimes\frac{\Id_{CD}}{d^2}\ket{\lambda_-}. \nonumber
\end{align}
Notice that the negative term $\lambda_-$ comes from the NPT-ness of the map  $\map{M_T}$, while the other term is due to the implementation of 
$\tilde{\map{T}_A}$ in the proposed experimental procedure. 
The expression above clearly shows that the operator $W_{\text{PPT}}$ can be 
regarded as a witness with respect to the set of PPT maps, as its expectation
value is always non-negative on this set. 
Therefore, if the expectation value
of the witness $W_{\text{PPT}}$ is negative, the map $\map{M}$ is guaranteed 
to be NPT.

Let us now assume that the map $\map{M}$ is unital \cite{nonunital}. The expectation 
value in Eq. \eqref{diff} then takes the simple form
\begin{equation}\label{exp_unit}
\Tr[ W_{\text{PPT}}C_{\map{M}\circ\tilde{\map{T}_A}}] = (1-p)\lambda_- + \frac{p}{d^4}.
\end{equation}
In this case the addition of the depolarized term that makes the map 
$\map{T}_A$ physically  implementable introduces only a constant shift in the 
expectation value of the witness. As a result, for any PPT unital map 
$\map{M}_{\text{PPT,unital}}$ we have
\begin{equation}\label{exp_ppt}
\Tr[ W_{\text{PPT}}C_{\map{M}_{\text{PPT,unital}}}\circ\tilde{\map{T}_A}] \geq  
\frac{p}{d^4}.
\end{equation}
Therefore, if we know a priori that the map $\map{M}$ to be detected is a 
unital map, then we are guaranteed that it is a NPT map 
whenever the expectation value of $W_{\text{PPT}}$ is smaller than ${p}/{d^4}$.

As an illustrative example we consider again the case of the CNOT gate. 
Here we want to detect such a gate as a NPT map by following the experimental 
procedure discussed above. It is straightforward to see that the Choi state 
$C_{\cnot_{\map{T}}}$  corresponding to the map 
$\cnot_{\map{T}}=\map{T}_A \circ\cnot\circ\map{T}_A$ has a single negative 
eigenvalue $\lambda_-=-1/2$. Since the CNOT is unital, from Eq. 
\eqref{exp_unit} it follows that 
$\Tr[W_\text{PPT}C_{\cnot\circ\tilde{\map{T}_A}}]=0$, and the gap with the 
bound provided by  Eq. \eqref{exp_ppt} ($\sim 0.055$ in this case) is 
then experimentally accessible.

\section{conclusion}
\label{conc}

In conclusion, we have presented an experimentally feasible method to detect
several sets of quantum channels. The proposed procedure works when some a priori knowledge on the quantum channel is available and is based on a link to detection 
methods for entanglement properties of multipartite quantum states via
witness operators. The method has been first explicitly illustrated in the 
simple case of entanglement breaking channels, and then presented to detect
separability properties of quantum channels. In particular, methods to reveal 
non separable random unitaries and non separable maps 
have been derived, showing also the possibility to detect the gap between the 
sets of SRUs and separable maps. 
This result highlights a richer separability structure of Choi operators that has no counterpart in the separability properties of ordinary entangled/separable states.
The present method can be also applied to 
other properties of quantum channels that rely on a convex structure and 
reflect on properties of the corresponding Choi states,
such as for example completely co-positive maps \cite{coco} or bi-entangling
operations introduced in Ref. \cite{virmani}.
The advantage over standard quantum 
process tomography
is that a much smaller number of measurement settings is needed in an 
experimental implementation. Finally, we want to point out that the proposed 
scheme can be implemented with current technology, for example in a quantum optical scheme \cite{exp}.

\section*{Acknowledgements}

We would like to thank Barbara Kraus for fruitful suggestions.


\begin{thebibliography}{99}
\bibitem{notabisep} 
Recall that states given by Eq. \eqref{ext} 
correspond to the generating points for SRU channels, which is actually a 
superset of the extremal points of the SRU set.
\bibitem{chi?} This can be proven by induction over the number of Kraus operators composing the quantum map and using the Cauchy-Schwartz inequality for the Hilbert-Schmidt inner product.
\bibitem{nota-3}
In order to achieve this, two general unitaries $U_A$, $U_B$ have been 
expressed as 
$U_A=\sum_{i=1}^{d^2}\alpha_i A_i$ and $U_B=\sum_{i=1}^{d^2}\beta_i B_i$,
with $\{\alpha_i\}_{i=1,...,d^2}$, $\{\beta_i\}_{i=1,...,d^2}$ complex 
coefficients fulfilling the unitary condition, and 
$\{A_i\}_{i=1,...,d^2}$, $\{B_i\}_{i=1,...,d^2}$ bases coming from the 
Schmidt decomposition of $Z_3$.
Numerical analysis shows that in this case it is not restrictive to consider two 
unitaries of the form $U_A=\alpha_1 A_1+\alpha_2 A_2$, 
$U_B=\beta_1 B_1+\beta_2 B_2$, which easily lead to a maximum overlap of 
$\alpha_{SRU}\sim 0.786$.
\bibitem{nonunital} Notice that the expectation value given by \eqref{diff} usually depends on the to-be-detected map $\map{M}$. Therefore, in order to find the optimal witness, one should minimize the last term of Eq. \eqref{diff} with respect to the set of all PPT maps, a hard task that drastically simplifies when the map $\map{M}$ is promised to be unital.
\end{thebibliography}
\end{document}